\documentclass[a4paper,11pt]{article}
\usepackage{pos}
\usepackage{setspace}

\title{The First Array-Wide Diffuse Flux Search for UHE Neutrinos with the Askaryan Radio Array}
\ShortTitle{The First Array-Wide Diffuse Flux Search for UHE Neutrinos with ARA}

\author[a]{Alan Salcedo-Gomez}
\author*[b,c]{Marco Stein Muzio}
\onbehalf{for the ARA Collaboration \\{\normalsize \normalfont(a complete list of authors can be found at the end of the proceedings)}\\}

\affiliation[a]{Dept. of Physics, Center for Cosmology and AstroParticle Physics, The Ohio State University,\\
Columbus, OH 43210, USA}
\affiliation[b]{Wisconsin IceCube Particle Astrophysics Center,\\
Madison, WI 53703, USA}
\affiliation[c]{Dept. of Physics, University of Wisconsin-Madison,\\
Madison, WI 53706, USA}

\emailAdd{salcedogomez.1@osu.edu}
\emailAdd{muzio@wisc.edu}

\abstract{
The Askaryan Radio Array (ARA) is an ultrahigh energy (UHE) neutrino detector at the South Pole that searches for impulsive broadband radio signals from neutrino-induced particle showers in glacial ice. ARA consists of five autonomous stations with receiving antennas deployed up to 200~m deep and has accumulated the largest livetime of any in-ice radio array. We present the first array-wide diffuse UHE neutrino search using data collected from January 2013 to December 2023, incorporating improved detector characterization and simulation, including data-driven noise and electronics models, revised antenna responses, and updated neutrino interaction and lepton propagation modeling, within a unified framework for event processing, background rejection, and cut optimization across independently operating stations. This search is expected either to identify the first UHE neutrino candidates observed by an in-ice radio detector or to set the most stringent UHE neutrino diffuse flux limits above a few EeV, while informing analysis strategies for under-construction and future radio arrays, such as RNO-G and IceCube-Gen2 Radio.
}

\FullConference{11th International Workshop on Acoustic and Radio EeV Neutrino Detection Activities (ARENA2026)\\
8-11 June 2026\\
Karlsruhe, Germany\\}

\begin{document}
\maketitle

\section{Introduction}
The origin of ultrahigh energy cosmic rays (UHECRs, $E_\mathrm{CR} \gtrsim 10^{18}$~eV) remains a central open problem in astroparticle physics. Significant observational progress over the past few decades has revealed that UHECRs are primarily of extragalactic origin and not purely protonic in composition~\cite{Roth:2025hqe,Kido:2025mey}. However, the direct identification of their sources is impeded by the deflections from the Galactic and extragalactic magnetic fields~\cite{Unger:2023lob,PierreAuger:2024hlp}. Moreover, their observable horizon is further limited at the highest energies to roughly 100~Mpc by interactions with the cosmic microwave background (CMB) and extragalactic background light (EBL)~\cite{Greisen1966,Zatsepin1966}.

Ultrahigh energy ($E_\nu \gtrsim 100$~PeV) neutrinos are produced by UHECR interactions both within their sources and during extragalactic propagation~\cite{Ackermann:2022rqc}. This makes UHE neutrinos a unique probe of UHECR sources, since they are unattenuated and undeflected in their propagation. Their detection can provide information of the physical conditions within UHECR sources and, through their arrival directions, potentially enable direct source identification.

Current constraints on the diffuse UHE neutrino flux~\cite{IceCubeCollaborationSS:2025jbi,ANITA:2019wyx} indicate that $\gtrsim 100$~km$^3$ detection volumes are required to observe these particles. Optical detectors such as IceCube and KM3NeT are difficult to scale, economically and logistically, to such volumes, motivating alternative detection techniques. UHE neutrinos interactions in the Antarctic ice produce Askaryan emission, a nanosecond-scale radio-wavelength impulse emitted by relativistic particle cascades in a dielectric medium~\cite{Askaryan:1961pfb}. Since radio waves propagate through ice with attenuation lengths of $\mathcal{O}(1)$~km and radio antennas are comparatively inexpensive to deploy, Askaryan detectors provide a scalable method to instrument the volumes required to discover the diffuse UHE neutrino flux.

\section{ARA Detector and Dataset}

The Askaryan Radio Array (ARA) is an in-ice radio detector located near the South Pole and the IceCube Neutrino Observatory. It consists of five autonomous stations, A1--A5, separated by roughly $2$~km. ARA searches for the radio-frequency (RF) impulses from UHE neutrino interactions in South Pole ice.

\begin{figure}[h!]
    \centering
    \includegraphics[width=0.49\linewidth]{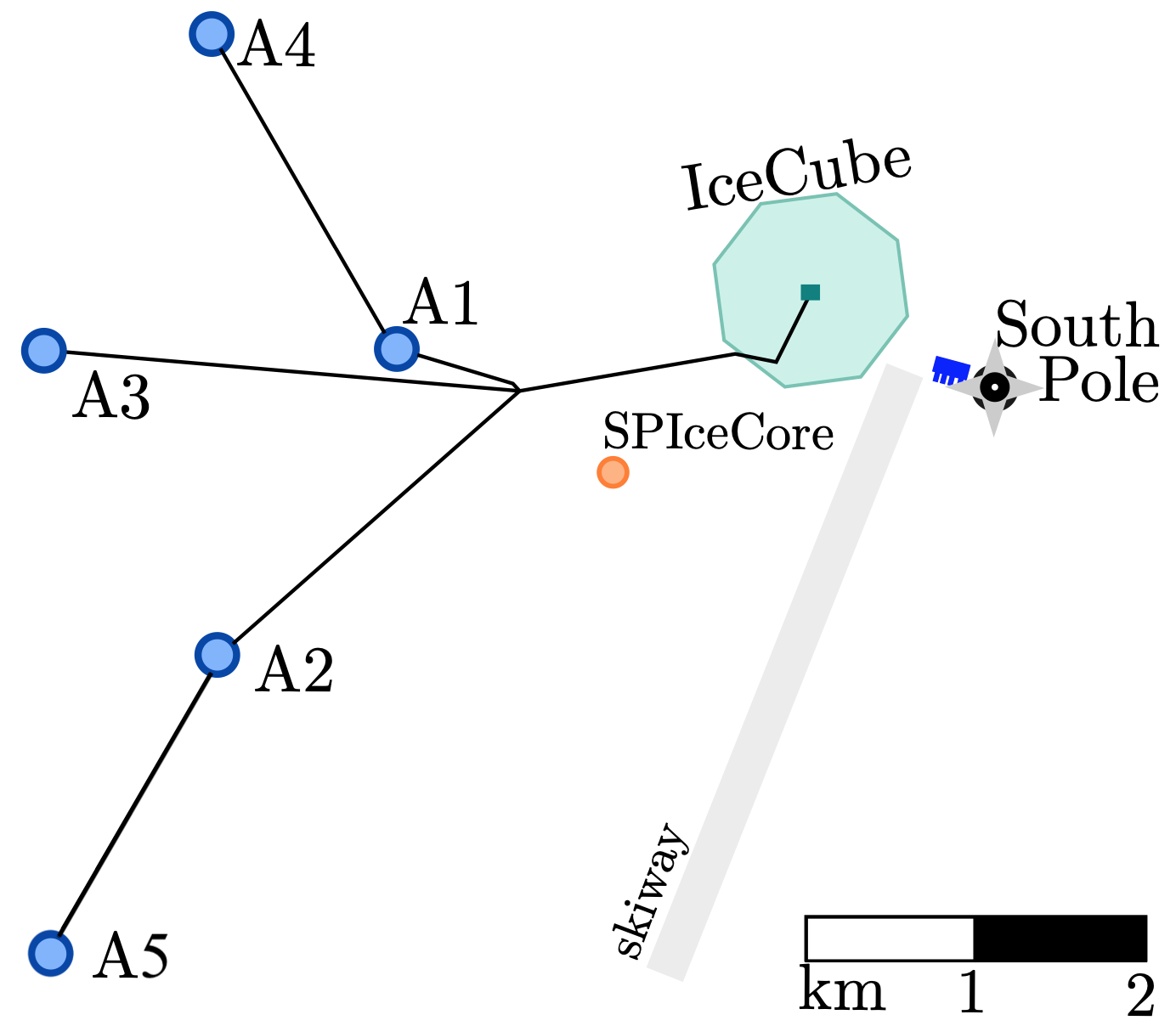}
    \includegraphics[width=0.49\linewidth]{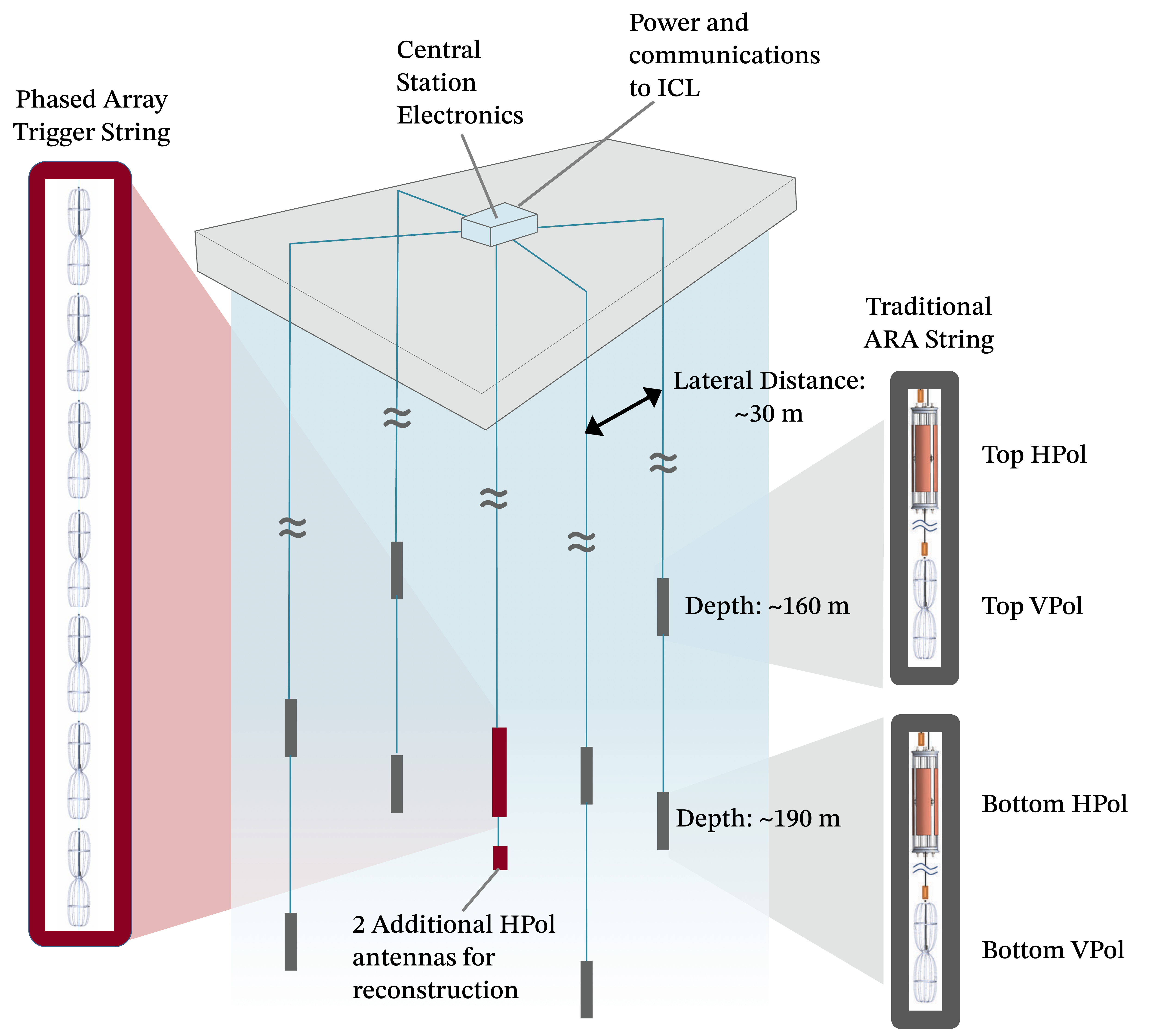}
    \caption{
        ARA's array layout (left) and station layout (right), shown for station A5. Each ARA station has four holes drilled into the ice and instrumented with radio antennas (shown in grey). A5 additionally has the phased array (PA) detector string (shown in red) deployed in the center of the four outer strings. A1--A3 have an installed lateral distance baseline of 14~m.
        \label{fig:ARA_station}
    }         
\end{figure}

A conventional ARA station consists of four instrumented boreholes separated by approximately $14$~m for A1--A3 and $30$~m for A4--A5, as illustrated for A5 in Fig.~\ref{fig:ARA_station}. The antennas extend to depths of approximately $90$~m at A1 and $190$--$200$~m at A2--A5. Each borehole contains four receiving antennas: two sensitive to vertical polarization (VPol) and two to horizontal polarization (HPol), for a total of 16 receiving antennas, or channels, per station. The antennas provide broadband sensitivity over the relevant frequencies for Askaryan emission between $150$--$850$~MHz. Stations additionally contain calibration pulsers, typically deployed on one or two dedicated strings, that are used for detector calibration and to monitor their stability.

Received signals are amplified by in-ice electronics, transmitted to the surface through radio-frequency-over-fiber cables, amplified again, and split into separate trigger and digitization paths. In the standard ARA trigger, a station-level trigger is issued when at least three antennas of the same polarization exceed a power threshold within an adjustable coincidence window. The thresholds maintain a station trigger rate near $5$~Hz, while software-forced and calibration triggers increase the total event rate to approximately $7$~Hz. Triggered events record approximately $500$~ns waveforms from all 16 channels.

Station A5 additionally hosts an independent phased array (PA) detector on a central string, also shown in Fig.~\ref{fig:ARA_station}, containing seven closely spaced VPol antennas and two HPol antennas. The PA trigger coherently sums time-delayed VPol waveforms given by predefined hypothesized arrival directions~\cite{Vieregg:2015baa,ARA:2018phased}. The amplitude of coherently-summed signals scales approximately as the number of antennas $N$, while uncorrelated thermal noise scales as $\sqrt{N}$, lowering the trigger threshold relative to the conventional ARA system. The PA has its own data acquisition system and is treated as a separate instrument in this analysis.

\begin{figure*}
    \centering\includegraphics[width=\textwidth]{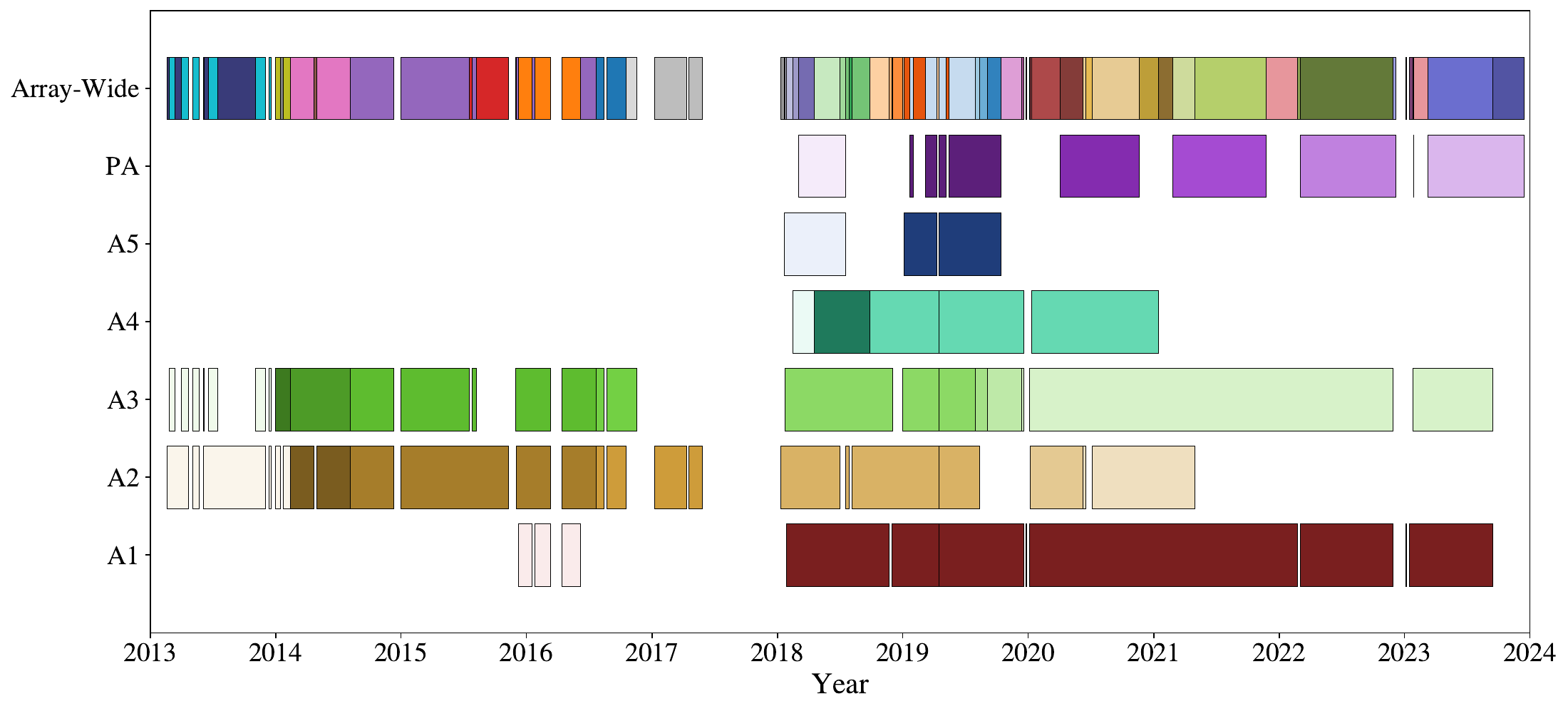}
    \caption{ARA detector configurations from January 2013 to December 2023. The lower rows show the station-level configurations of A1--A5 and the PA, while the top row shows the corresponding array-wide configurations. Repeated colors in a row denote the same configuration.}
    \label{fig:arraywide_configs}
\end{figure*}

Over the January 2013--December 2023 analysis period, changes in station operating conditions divide each detector dataset into station-level configurations, corresponding to periods of approximately stable detector response. Simultaneously active station-level configurations define the array-wide configurations used throughout the analysis, as shown in Fig.~\ref{fig:arraywide_configs}. After data-quality selections, these configurations provide a total of 10.6~years of array-wide livetime.

\section{Event Selection}

ARA triggers are dominated by thermal noise, calibration pulsers, anthropogenic radio emissions, and cosmic ray air showers. Event selection is designed to suppress these backgrounds while retaining neutrino-induced radio signals across the array.

The selection is performed in four stages. First, data quality cuts remove periods of unreliable operation and events with inconclusive interferometric reconstruction. Second, non-thermal event selections reject backgrounds from calibration pulsers, anthropogenic activity, and cosmic rays. Third, the final thermal and surface-origin cut thresholds are optimized simultaneously for discovery potential. Finally, an array-wide time-isolation requirement removes remaining anthropogenic backgrounds. The selections and background models were developed using a $10\%$~sample of the data while simulated neutrinos were used to evaluate signal efficiency and optimize the final selection thresholds for $5\sigma$~discovery potential.

\subsection{Data Quality Cuts}

Data quality selections remove periods affected by detector maintenance, calibration campaigns, or high anthropogenic activity. Contaminated runs are identified using collaboration logbooks and by searching for non-thermal populations in the distributions of interferometric reconstruction correlations. The first minute of each run is also excluded while the trigger-threshold servo stabilizes.
Events containing detector-generated emissions are removed by identifying anomalous power concentrated on a single detector string or channel~\cite{ARA:2019wcf}. Events with non-thermal waveform characteristics but poor interferometric reconstruction are also rejected.

\subsection{Non-thermal Event Cuts}

Several selections are applied to remove non-thermal backgrounds. Directional cuts remove events associated with known sources, particularly local calibration pulsers. These events are normally identified by event tags, while untagged events are rejected using their reconstructed directions.
Transient anthropogenic backgrounds are removed using a spatiotemporal clustering cut, which identifies signal-like events that reconstruct to a common region of the sky within a limited time interval.
The remaining non-thermal backgrounds are dominated by anthropogenic and cosmic ray events originating near or above the ice surface. These are rejected using interferometric surface cuts that remove events whose primary reconstruction, or significant secondary reconstruction solutions, are consistent with a surface origin.

\subsection{Optimized Thermal and Surface Event Cuts}

After the non-thermal selections, the remaining data are dominated by thermal-like events. These events are separated from simulated neutrinos using a linear discriminant analysis (LDA) trained for each station-level detector configuration. The LDA combines several analysis variables into a single score, with neutrino-like events preferentially occupying the high-score region. The tail of the thermal-like population in the $10\%$~sample is parametrized to estimate the expected background leakage into the signal region.

The final LDA and surface event cut thresholds are optimized simultaneously for $5\sigma$ discovery potential over the full array livetime. The optimization spans 150 parameters: 144 LDA thresholds, one for each active station within an array-wide configuration, and six detector-level surface thresholds. The expected thermal and surface background leakages are evaluated as these thresholds are varied.
After optimization, the expected thermal and surface backgrounds are $8.65^{+0.44}_{-0.28}\times10^{-2}$ and $4.50^{+1.01}_{-0.85}\times10^{-2}$ events, respectively. Including a small contribution from untagged calibration pulsers gives a total expected background of $0.13\pm0.01$ events over the full analysis livetime.

\subsection{Array-Wide Time Isolation Cut}
A final array-wide selection is applied to suppress residual anthropogenic backgrounds. Since neutrinos from a diffuse flux are expected to be widely separated in time, surviving events are required to be isolated from other candidates by at least $5.8$~hours. For an optimistic UHE neutrino flux consistent with current experimental limits~\cite{ANITA:2019wyx,IceCubeCollaborationSS:2025jbi}, we conservatively expect up to 15 neutrino events over the $10.6$~year analysis livetime. In this case, $99.9\%$ of consecutive neutrino events are expected to be separated by at least $5.8$~hours. Events detected at different stations within $5$~s are flagged for inspection before removal to preserve potential multi-station neutrino events~\cite{ARA:2026wjc}.

\section{Simulations \& Projected Sensitivity}

ARA's sensitivity is calculated in three steps: neutrino event generation, detector simulation, and array-wide sensitivity calculation. Neutrino interactions are generated with \texttt{NuLeptonSim}~\cite{Cummings:2023iuw}, which propagates primary neutrinos and their secondary particles through the Earth. Interactions produced near the detector are stored in common event libraries and passed to \texttt{AraSim}, which simulates the Askaryan emission, its propagation through the ice, the detector response, and the trigger. To reduce computational cost, each station and detector configuration is simulated separately and then combined using the corresponding array-wide configuration and livetime. A detailed description and validation of the simulation framework is presented in a separate publication~\cite{ARA:2026wjc}.

\subsection{Neutrino Event Generation}

\texttt{NuLeptonSim} generates neutrinos with trajectories intersecting a fiducial volume surrounding the array and propagates them through the Earth toward the detector. Neutrino interactions are sampled using the neutrino-nucleon cross section model of~\cite{ctw}, while cascade and secondary particle energies are determined using inelasticity tables from the CTEQ5 parton distribution~\cite{cteq5}. Secondary particles are propagated until they leave the fiducial volume or fall below an energy threshold
$E_\mathrm{th}=\min(10~\mathrm{PeV},E_\nu/10)$.
Interactions inside the fiducial volume are recorded with their energy, position, direction, interaction type, and a primary-neutrino identification number.

Multiple interactions from the same primary neutrino can therefore be incorporated, including those produced from secondary muons and taus. These interactions form the event libraries passed to the detector simulation. Only neutrinos producing interactions inside the fiducial volume are retained, so Monte Carlo weights are applied when calculating the detector acceptance to correct for this sampling bias.

\subsection{Detector Simulation}

The event libraries are passed to \texttt{AraSim}~\cite{ARA:2026wjc}, which generates the Askaryan radio emission from each particle cascade and propagates the signal to the detector. Ray tracing through Antarctic ice accounts for the depth-dependent index of refraction and signal attenuation. The resulting electric fields are converted to antenna voltages and passed through models of the antenna and electronics response. Thermal noise is then added to produce simulated detector waveforms.

Interactions from the same primary neutrino are combined into a single waveform with the appropriate relative arrival times, allowing the simulation to model events containing multiple radio pulses from secondary particle interactions. The detector trigger is then applied to the full waveform using the appropriate trigger for each station.

For this analysis, \texttt{AraSim} includes updated data-driven models of the detector response. Antenna responses are based on anechoic chamber measurements, while the system gain and thermal-noise models are derived from in-situ detector data for each channel and detector configuration. These updates account for changes in detector response over the full data-taking period and improve agreement between simulated and measured thermal noise events.

\subsection{Projected Sensitivity}

The station-level simulations are combined to calculate the response of the full array. For each array-wide configuration, triggers from the active stations are merged so that a primary neutrino triggering multiple stations is counted only once. The acceptance is then calculated from the weighted fraction of simulated primary neutrinos that trigger the array.

The total exposure is obtained by combining the acceptance of each array-wide configuration with its corresponding livetime and analysis efficiency. Assuming an equal flavor ratio of $\nu_e:\nu_\mu:\nu_\tau=1:1:1$ and equal neutrino and antineutrino fluxes, the flavor-averaged exposure is
\begin{align}
\mathcal{E}(E) = \sum_c \mathcal{A}_c(E) T_c \varepsilon_c(E)~,
\end{align}
where $\mathcal{A}_c(E)$, $T_c$, and $\varepsilon_c(E)$ are the acceptance, livetime, and analysis efficiency of array-wide configuration $c$, respectively.

The projected all-flavor diffuse neutrino sensitivity is calculated from this exposure in decade-wide energy bins. For zero observed events, the $90\%$ confidence level differential flux limit is
\begin{align}
\phi^{\mathrm{UL}}_{\nu+\bar{\nu}}(E_0) = \frac{N_{\mathrm{UL}}}{\ln(10) E_0 \mathcal{E}(E_0)}~,
\end{align}
where $N_{\mathrm{UL}}=2.44$ is the Feldman--Cousins upper limit for a background-free search~\cite{feldman-cousins}, and $E_0$ is the logarithmic center of the energy bin with spectrum $E^{-1}$.

\section{Results}

\begin{figure}[h!]
    \centering
    \includegraphics[width=0.65\linewidth]{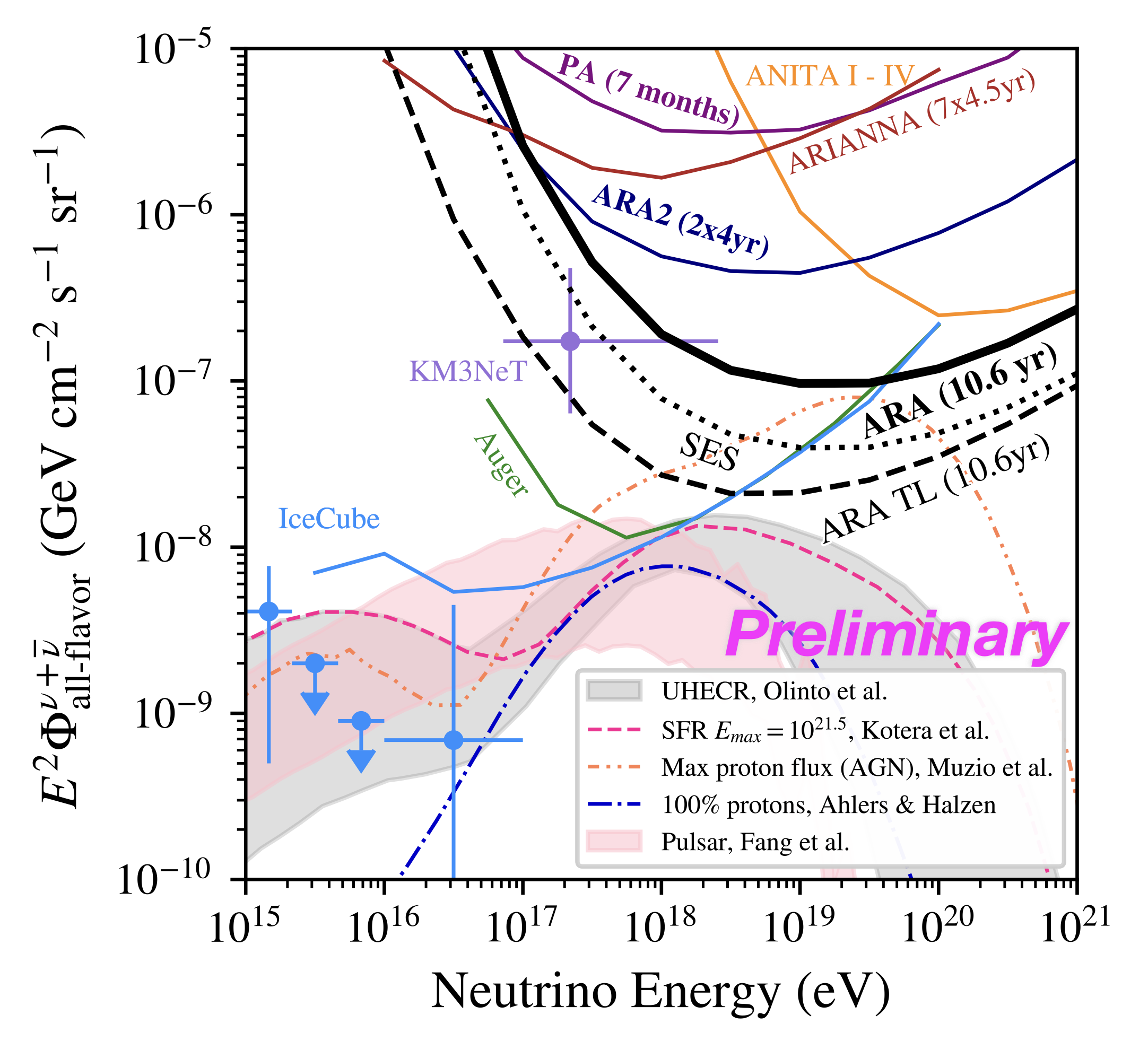}
    \caption{
        Upper limit on the diffuse UHE neutrino flux set by this analysis (bold black line). The analysis-level single event sensitivity (SES) and trigger-level (TL) upper limit are shown for reference. Flux measurements and constraints from other experiments and previous ARA searches, along with model predictions, are also shown. 
        \label{fig:limit}
    }         
\end{figure}

A single event on the PA passed the event selection prior to the time isolation cut. Upon investigation, this event was determined to be a detector background and removed via an \textit{a posteriori} cut. The final search therefore yielded zero observed events, with an expected background of $0.13\pm 0.01$~events. We consequently place an upper limit on the diffuse UHE neutrino flux, shown in Fig.~\ref{fig:limit}. This result will provide the strongest limit from any radio detector array and the leading constraint above ${\sim}40$~EeV. 

\section{Conclusion}

We have conducted the first array-wide search for UHE neutrinos with a radio detector array using the full ARA dataset. This analysis was developed on a $10\%$~sample of more than a decade of data and combines station-level background suppression, array-wide event selection optimized for discovery potential, and updated detector and neutrino simulations. No events survived the final selection in the $100\%$ dataset, and therefore we expect to set the leading upper limit on the diffuse UHE neutrino flux above ${\sim}40$~EeV. This analysis demonstrates the feasibility of array-wide neutrino searches and provides a foundation for future searches with larger radio detector arrays, including the Radio Neutrino Observatory in Greenland (RNO-G)~\cite{RNO-G:2020rmc} and IceCube-Gen2 Radio~\cite{IceCubeGen2_TDR}.

\begingroup
\setstretch{0.5}
\setlength{\bibsep}{1.0pt}
\bibliographystyle{JHEP}
\bibliography{references}

@article{Askaryan:1961pfb,
    author = "Askaryan, G. A.",
    title = "{Excess negative charge of an electron-photon shower and its coherent radio emission}",
    journal = "Sov. Phys. JETP",
    volume = "14",
    pages = "441--443",
    year = "1962"
}

@article{ARA:2018phased,
    author = "Allison, P. and others",
    title = "{Design and performance of an interferometric trigger array for radio detection of high-energy neutrinos}",
    eprint = "1809.04573",
    archivePrefix = "arXiv",
    primaryClass = "astro-ph.IM",
    doi = "10.1016/j.nima.2019.01.067",
    journal = "Nucl. Instrum. Meth. A",
    volume = "930",
    pages = "112--125",
    year = "2019"
}

@article{Vieregg:2015baa,
    author = "Vieregg, A. G. and Bechtol, K. and Romero-Wolf, A.",
    title = "{A Technique for Detection of PeV Neutrinos Using a Phased Radio Array}",
    eprint = "1504.08006",
    archivePrefix = "arXiv",
    primaryClass = "astro-ph.IM",
    doi = "10.1088/1475-7516/2016/02/005",
    journal = "JCAP",
    volume = "02",
    pages = "005",
    year = "2016"
}

@article{ARA:2019wcf,
    author = "Allison, P. and others",
    collaboration = "ARA",
    title = "{Constraints on the diffuse flux of ultrahigh energy neutrinos from four years of Askaryan Radio Array data in two stations}",
    eprint = "1912.00987",
    archivePrefix = "arXiv",
    primaryClass = "astro-ph.HE",
    doi = "10.1103/PhysRevD.102.043021",
    journal = "Phys. Rev. D",
    volume = "102",
    number = "4",
    pages = "043021",
    year = "2020"
}

@article{ARA:2026wjc,
    author = "Alden, N. and others",
    collaboration = "ARA",
    title = "{Sensitivity of the As-Built Askaryan Radio Array to Ultra-High Energy Neutrinos}",
    eprint = "2605.04268",
    archivePrefix = "arXiv",
    primaryClass = "astro-ph.HE",
    month = "5",
    year = "2026"
}

@article{ANITA:2019wyx,
    author = "Gorham, P. W. and others",
    collaboration = "ANITA",
    title = "{Constraints on the ultrahigh-energy cosmic neutrino flux from the fourth flight of ANITA}",
    eprint = "1902.04005",
    archivePrefix = "arXiv",
    primaryClass = "astro-ph.HE",
    doi = "10.1103/PhysRevD.99.122001",
    journal = "Phys. Rev. D",
    volume = "99",
    number = "12",
    pages = "122001",
    year = "2019"
}

@article{IceCubeCollaborationSS:2025jbi,
    author = "Abbasi, R. and others",
    collaboration = "IceCube",
    title = "{Search for Extremely-High-Energy Neutrinos and First Constraints on the Ultrahigh-Energy Cosmic-Ray Proton Fraction with IceCube}",
    eprint = "2502.01963",
    archivePrefix = "arXiv",
    primaryClass = "astro-ph.HE",
    doi = "10.1103/PhysRevLett.135.031001",
    journal = "Phys. Rev. Lett.",
    volume = "135",
    number = "3",
    pages = "031001",
    year = "2025"
}

@article{Cummings:2023iuw,
    author = "Cummings, Austin and Krebs, Ryan and Wissel, Stephanie and Alvarez-Mu{\~n}iz, Jaime and Carvalho, Jr., Washington R. and Romero-Wolf, Andr{\'e}s and Schoorlemmer, Harm and Zas, Enrique",
    title = "{Secondary lepton production, propagation, and interactions}",
    eprint = "2311.03646",
    archivePrefix = "arXiv",
    primaryClass = "hep-ph",
    doi = "10.1103/PhysRevD.111.023012",
    journal = "Phys. Rev. D",
    volume = "111",
    number = "2",
    pages = "023012",
    year = "2025"
}

@article{ctw,
    author = "Connolly, Amy and Thorne, Robert S. and Waters, David",
    title = "{Calculation of High Energy Neutrino-Nucleon Cross Sections and Uncertainties Using the MSTW Parton Distribution Functions and Implications for Future Experiments}",
    eprint = "1102.0691",
    archivePrefix = "arXiv",
    primaryClass = "hep-ph",
    doi = "10.1103/PhysRevD.83.113009",
    journal = "Phys. Rev. D",
    volume = "83",
    pages = "113009",
    year = "2011"
}

@article{feldman-cousins,
    author = "Feldman, Gary J. and Cousins, Robert D.",
    title = "{A Unified approach to the classical statistical analysis of small signals}",
    eprint = "physics/9711021",
    archivePrefix = "arXiv",
    reportNumber = "HUTP-97-A096",
    doi = "10.1103/PhysRevD.57.3873",
    journal = "Phys. Rev. D",
    volume = "57",
    pages = "3873--3889",
    year = "1998"
}

@article{Roth:2025hqe,
    author = "Roth, Markus",
    collaboration = "Pierre Auger",
    title = "{Exploring the Ultra-High-Energy Universe: Highlights from the Pierre Auger Observatory}",
    doi = "10.22323/1.501.1402",
    journal = "PoS",
    volume = "ICRC2025",
    pages = "1402",
    year = "2025"
}

@article{Kido:2025mey,
    author = "Kido, Eiji",
    collaboration = "Telescope Array",
    title = "{Recent Progress of the Telescope Array Experiment}",
    doi = "10.22323/1.501.1410",
    journal = "PoS",
    volume = "ICRC2025",
    pages = "1410",
    year = "2025"
}

@article{Unger:2023lob,
    author = "Unger, Michael and Farrar, Glennys R.",
    title = "{The Coherent Magnetic Field of the Milky Way}",
    eprint = "2311.12120",
    archivePrefix = "arXiv",
    primaryClass = "astro-ph.GA",
    doi = "10.3847/1538-4357/ad4a54",
    journal = "Astrophys. J.",
    volume = "970",
    number = "1",
    pages = "95",
    year = "2024"
}

@article{PierreAuger:2024hlp,
    author = "Abdul Halim, A. and others",
    collaboration = "Pierre Auger",
    title = "{Impact of the magnetic horizon on the interpretation of the Pierre Auger Observatory spectrum and composition data}",
    eprint = "2404.03533",
    archivePrefix = "arXiv",
    primaryClass = "astro-ph.HE",
    reportNumber = "FERMILAB-PUB-24-0144-CSAID-PPD-TD-V",
    doi = "10.1088/1475-7516/2024/07/094",
    journal = "JCAP",
    volume = "07",
    pages = "094",
    year = "2024"
}

@article{Greisen1966,
    author = "Greisen, Kenneth",
    title = "{End to the Cosmic-Ray Spectrum?}",
    doi = "10.1103/PhysRevLett.16.748",
    journal = "Phys. Rev. Lett.",
    volume = "16",
    pages = "748--750",
    year = "1966"
}

@article{Zatsepin1966,
    author = "Zatsepin, G. T. and Kuzmin, V. A.",
    title = "{Upper limit of the spectrum of cosmic rays}",
    journal = "JETP Lett.",
    volume = "4",
    pages = "78--80",
    year = "1966"
}

@article{Ackermann:2022rqc,
    author = "Ackermann, M. and others",
    title = "{High-energy and ultra-high-energy neutrinos: A Snowmass white paper}",
    eprint = "2203.08096",
    archivePrefix = "arXiv",
    primaryClass = "hep-ph",
    doi = "10.1016/j.jheap.2022.08.001",
    journal = "J. High Energy Astrophys.",
    volume = "36",
    pages = "55--110",
    year = "2022"
}

@article{RNO-G:2020rmc,
    author = "Aguilar, J. A. and others",
    collaboration = "RNO-G",
    title = "{Design and performance of the Radio Neutrino Observatory Greenland (RNO-G)}",
    eprint = "2010.12279",
    archivePrefix = "arXiv",
    primaryClass = "astro-ph.IM",
    doi = "10.1088/1748-0221/16/03/P03032",
    journal = "JINST",
    volume = "16",
    number = "03",
    pages = "P03032",
    year = "2021"
}

@article{IceCubeGen2_TDR,
    author = "Aartsen, M. G. and others",
    collaboration = "IceCube-Gen2",
    title = "{IceCube-Gen2: the next-generation neutrino observatory for the South Pole}",
    eprint = "2008.04323",
    archivePrefix = "arXiv",
    primaryClass = "astro-ph.HE",
    doi = "10.1088/1361-6471/abbd48",
    journal = "J. Phys. G",
    volume = "48",
    number = "6",
    pages = "060501",
    year = "2021"
}

@article{cteq5,
    author = "Kuhlmann, S.",
    editor = "Blumlein, J. and Riemann, T.",
    title = "{CTEQ5 parton distributions and ongoing studies}",
    reportNumber = "ANL-HEP-CP-99-95",
    doi = "10.1016/S0920-5632(99)00648-9",
    journal = "Nucl. Phys. B Proc. Suppl.",
    volume = "79",
    pages = "108--110",
    year = "1999"
}
\endgroup

\clearpage

\section*{Full Author List: ARA Collaboration (August 24, 2026)}

\noindent
N.~Alden\textsuperscript{1}, 
S.~Ali\textsuperscript{2}, 
P.~Allison\textsuperscript{3}, 
J.J.~Beatty\textsuperscript{3}, 
D.Z.~Besson\textsuperscript{2}, 
A.~Bishop\textsuperscript{4}, 
P.~Chen\textsuperscript{5}, 
Y.C.~Chen\textsuperscript{5}, 
Y.-C.~Chen\textsuperscript{5}, 
S.~Chiche\textsuperscript{6}, 
B.A.~Clark\textsuperscript{7}, 
A.~Connolly\textsuperscript{3}, 
K.~Couberly\textsuperscript{2}, 
L.~Cremonesi\textsuperscript{8}, 
A.~Cummings\textsuperscript{9,10,11}, 
P.~Dasgupta\textsuperscript{3}, 
R.~Debolt\textsuperscript{3}, 
S.~de~Kockere\textsuperscript{12}, 
K.D.~de~Vries\textsuperscript{12}, 
C.~Deaconu\textsuperscript{1}, 
M.A.~DuVernois\textsuperscript{4}, 
J.~Flaherty\textsuperscript{3}, 
E.~Friedman\textsuperscript{7}, 
R.~Gaior\textsuperscript{13}, 
P.~Giri\textsuperscript{14}, 
J.~Hanson\textsuperscript{15}, 
N.~Harty\textsuperscript{16}, 
K.D.~Hoffman\textsuperscript{7}, 
M.-H.~Huang\textsuperscript{5,17}, 
K.~Hughes\textsuperscript{3}, 
A.~Ishihara\textsuperscript{13}, 
A.~Karle\textsuperscript{4}, 
J.L.~Kelley\textsuperscript{4}, 
K.-C.~Kim\textsuperscript{7}, 
M.-C.~Kim\textsuperscript{13}, 
I.~Kravchenko\textsuperscript{14}, 
R.~Krebs\textsuperscript{9,10}, 
C.Y.~Kuo\textsuperscript{5}, 
U.A.~Latif\textsuperscript{12}, 
C.H.~Liu\textsuperscript{14}, 
T.C.~Liu\textsuperscript{5,18}, 
W.~Luszczak\textsuperscript{3}, 
A.~Machtay\textsuperscript{3}, 
M.S.~Muzio\textsuperscript{4,9,10,11}, 
J.~Nam\textsuperscript{5}, 
R.J.~Nichol\textsuperscript{8}, 
A.~Novikov\textsuperscript{16}, 
A.~Nozdrina\textsuperscript{3}, 
E.~Oberla\textsuperscript{1}, 
C.W.~Pai\textsuperscript{5}, 
Y.~Pan\textsuperscript{16}, 
C.~Pfendner\textsuperscript{19}, 
N.~Punsuebsay\textsuperscript{16}, 
J.~Roth\textsuperscript{16}, 
A.~Salcedo-Gomez\textsuperscript{3}, 
D.~Seckel\textsuperscript{16}, 
M.F.H.~Seikh\textsuperscript{2}, 
Y.-S.~Shiao\textsuperscript{5,20}, 
J.~Stethem\textsuperscript{3}, 
S.C.~Su\textsuperscript{5}, 
S.~Toscano\textsuperscript{6}, 
J.~Torres\textsuperscript{3}, 
J.~Touart\textsuperscript{7}, 
N.~van~Eijndhoven\textsuperscript{12}, 
A.~Vieregg\textsuperscript{1}, 
M.~Vilarino~Fostier\textsuperscript{6}, 
M.-Z.~Wang\textsuperscript{5}, 
S.-H.~Wang\textsuperscript{5}, 
P.~Windischhofer\textsuperscript{1}, 
S.A.~Wissel\textsuperscript{9,10,11}, 
C.~Xie\textsuperscript{8}, 
S.~Yoshida\textsuperscript{13}, 
R.~Young\textsuperscript{2}
\\
\\
\textsuperscript{1} Dept. of Physics, Dept. of Astronomy and Astrophysics, Enrico Fermi Institute, Kavli Institute for Cosmological Physics, University of Chicago, Chicago, IL 60637\\
\textsuperscript{2} Dept. of Physics and Astronomy, University of Kansas, Lawrence, KS 66045\\
\textsuperscript{3} Dept. of Physics, Center for Cosmology and AstroParticle Physics, The Ohio State University, Columbus, OH 43210\\
\textsuperscript{4} Dept. of Physics, Wisconsin IceCube Particle Astrophysics Center, University of Wisconsin-Madison, Madison,  WI 53706\\
\textsuperscript{5} Dept. of Physics, Graduate Institute of Astrophysics, Leung Center for Cosmology and Particle Astrophysics, National Taiwan University, Taipei, Taiwan\\
\textsuperscript{6} Universite Libre de Bruxelles, Science Faculty CP230, B-1050 Brussels, Belgium\\
\textsuperscript{7} Dept. of Physics, University of Maryland, College Park, MD 20742\\
\textsuperscript{8} Dept. of Physics and Astronomy, University College London, London, United Kingdom\\
\textsuperscript{9} Center for Multi-Messenger Astrophysics, Institute for Gravitation and the Cosmos, Pennsylvania State University, University Park, PA 16802\\
\textsuperscript{10} Dept. of Physics, Pennsylvania State University, University Park, PA 16802\\
\textsuperscript{11} Dept. of Astronomy and Astrophysics, Pennsylvania State University, University Park, PA 16802\\
\textsuperscript{12} Vrije Universiteit Brussel, Brussels, Belgium\\
\textsuperscript{13} International Center for Hadron Astrophysics, Chiba University, Chiba 263-8522 Japan\\
\textsuperscript{14} Dept. of Physics and Astronomy, University of Nebraska, Lincoln, Nebraska 68588\\
\textsuperscript{15} Dept. Physics and Astronomy, Whittier College, Whittier, CA 90602\\
\textsuperscript{16} Dept. of Physics, University of Delaware, Newark, DE 19716\\
\textsuperscript{17} Dept. of Energy Engineering, National United University, Miaoli, Taiwan\\
\textsuperscript{18} Dept. of Applied Physics, National Pingtung University, Pingtung City, Pingtung County 900393, Taiwan\\
\textsuperscript{19} Dept. of Physics and Astronomy, Denison University, Granville, Ohio 43023\\
\textsuperscript{20} National Nano Device Laboratories, Hsinchu 300, Taiwan\\

\noindent
The ARA Collaboration is grateful for support from the National Science Foundation through Award No.~2013134 and~2310095.
The ARA Collaboration designed, constructed, and now operates the ARA detectors. 
We would like to thank IceCube, and specifically the winterovers, for the support in operating the detector. 
Data processing and calibration, Monte Carlo simulations of the detector and of theoretical models, and data analyses were performed by a large number
of collaboration members, who also discussed and approved the scientific results presented here. 
We are thankful to Antarctic Support Contractor staff, a Leidos unit for field support and enabling our work on the harshest continent. 
We thank the National Science Foundation (NSF) Office of Polar Programs and Physics Division for funding support. 
We further thank the Taiwan National Science Council's Vanguard Program NSC 92-2628-M-002-09 and the Belgian F.R.S.-FNRS and FWO.
K.~Hughes thanks the NSF for support through the Graduate Research Fellowship Program Award No.~1746045. 
A.~Connolly thanks the NSF for Awards No.~1806923 and No.~2209588 and also acknowledges the Ohio Supercomputer Center. 
S.~A.~Wissel thanks the NSF for support through CAREER Award No.~2033500.
A.~Vieregg, C.~Deaconu, N.~Alden, and P.~Windischhofer thank the NSF for Award No.~2411662 and the Research Computing Center at the University of Chicago
for computing resources.
R.~Nichol thanks the Leverhulme Trust for their support. 
K.D.~de~Vries is supported by European Research Council under the European Union's Horizon research and innovation program (Grant Agreement 763 No.~805486). 
D.~Besson, I.~Kravchenko, and D.~Seckel thank the NSF for support through the IceCube EPSCoR Initiative (Award ID No.~2019597). 
M.S.~Muzio thanks the NSF for support through the MPS-Ascend Postdoctoral Fellowship under Award No.~2138121. 
A.~Bishop thanks the Belgian American Education Foundation for their Graduate Fellowship support.

\end{document}